\documentclass[%
reprint,
 amsmath,amssymb,
 aps,
showpacs,showkeys,]{revtex4-1}

\usepackage{graphicx}% Include figure files
\usepackage{dcolumn}% Align table columns on decimal point
\usepackage{bm}% bold math

\begin{document}

\title{Parametrization of unitary matrices, parametrizing the symmetric $S$ matrix and the $NN$ $S$ matrix above the inelastic threshold}

\author{N. A. Khokhlov}
\email{nikolakhokhlov@yandex.ru; nakhokhlov@gmail.com}
\affiliation{%
Peter the Great St. Petersburg Polytechnic University, St. Petersburg, Russia.
}%
\date{\today}% 

\begin{abstract}
None of the $S$-matrix representations used in analyses of nucleon-nucleon ($NN$) scattering above the inelastic threshold is implicitly unitary, general, and connected to low-energy (Stapp) parameters, and hence correspondingly related to the physical properties of the scattering. Such a representation is also needed to solve the corresponding inverse scattering problem. I present an $S$-matrix representation which satisfies these requirements. As a by-product, I derive a parametrization of an $n$-dimensional unitary matrix.
\end{abstract}
\maketitle
\section{\label{sec:intro}INTRODUCTION}
There are a number of $S$-matrix models~\cite{Blatt1952,Stapp1957,MacGregor1968,Hoshizaki1968} used in phase-shift analyses. Not all of them are equivalent. Since it is based on a diagonalization of a unitary symmetric matrix by rotation matrices, only the parametrization by Blatt and Biedenharn~\cite{Blatt1952} is easily generalized to the multichannel case with an arbitrary number of channels. Unfortunately, this parametrization presents certain difficulties, which have led, for the most part, to the use of the parametrization of Ref.~\cite{Stapp1957} in the analysis of low-energy $NN$ elastic scattering.
In the 1980s, a number of papers~\cite{Bryan1981,Bryan1984,Arndt1982,Arndt1983,Sprung1982,Sprung1985} appeared discussing a possible parametrization of the $S$-matrix that may be used for the description of $NN$ scattering above the inelastic threshold. In these articles, two possible parametrizations of the $S$-matrix for two coupled channels were proposed, accounting for the coupling of these channels to other open channels. 
 Neither parametrization is entirely satisfactory for the stated aims, as noted in those works. Furthermore, they do not constitute a general extension of the parametrization of Ref.~\cite{Stapp1957}.
 
 To construct a parametrization for the $S$-matrix, I had to develop a parametrization for general unitary matrices. Although several solutions to this problem already exist \cite{Murnaghan,Dita1,Dita2,Jalskog}, the task of constructing new parametrizations for unitary matrices remains relevant, as a specific parametrization may prove more convenient than others for certain applications \cite{Jalskog}.
 
In this paper I generalize the Stapp parametrization to the case of $S$ matrix of arbitrary dimension. 
  The Stapp parameters are defined by equation 
  	\begin{eqnarray}
  S=
  	\left[\begin{array}{cc}
  		\cos 2{\bar \epsilon} {\text e}^{2\imath {\bar \delta}_1 } & \imath \sin 2{\bar \epsilon} {\text e}^{\imath \left({\bar \delta}_1 +{\bar \delta}_2\right)} 			\\
  		%%%%%%%%%%%%%%%%%%%%%%%	
  	\imath \sin 2{\bar \epsilon} {\text e}^{\imath \left({\bar \delta}_1 +{\bar \delta}_2\right)} 		   & \cos2{\bar \epsilon} {\text e}^{2\imath {\bar \delta}_2 }
  	\end{array}\right]\nonumber \\\label{Stapp2} 
  	\equiv
  		\left[\begin{array}{cc}
  		{\text e}^{\imath {\bar \delta}_1 } & 0 	\\
  		%%%%%%%%%%%%%%%%%%%%%%%	
  		0	   & {\text e}^{\imath {\bar \delta}_2 }
  	\end{array}\right]
  	  	 	\left[\begin{array}{cc}
  		\cos 2{\bar \epsilon} & \imath \sin 2{\bar \epsilon} 	\\
  		%%%%%%%%%%%%%%%%%%%%%%%	
  		\imath \sin 2{\bar \epsilon}	   & \cos2{\bar \epsilon} 
  	\end{array}\right]
  		\left[\begin{array}{cc}
  		{\text e}^{\imath {\bar \delta}_1 } & 0 	\\
  		%%%%%%%%%%%%%%%%%%%%%%%	
  		0	   & {\text e}^{\imath {\bar \delta}_2 }
  	\end{array}\right].
  \end{eqnarray}
 The proposed parametrization, which generalizes Eq.~(\ref{Stapp2}) to an arbitrary number $n$ of coupled channels, 
 also reduces to the $n=2$ case when the coupling between all channels—except between any two chosen channels—is switched off.

\section{Building blocks}
The parametrization is based on factorizing a unitary matrix into the following unitary factors:
\begin{itemize}
	\item diagonal $n$-dimensional phase matrices,
	\begin{equation}
		\Delta^{(n)}(\delta_{1},\delta_{2},\dots,\delta_{n})=
		diag\left({\mathrm e}^{\imath \delta_1},{\mathrm e}^{\imath \delta_2},\dots,{\mathrm e}^{\imath \delta_n}\right);
		\label{Delta_matrix}
	\end{equation}
	\item rotation matrices,  $\Phi^{(n)}_{kj}$ denotes the $n$-dimensional matrix defined as follows:
	all diagonal elements are $1$ save the $k$th and $j$th which are each $C_{jk}=\cos (\phi_{jk} )$, and
	all  nondiagonal elements are zero save the elements in the $k$th column and $j$th row, which is  $S_{jk}= \sin(\phi_{jk} )$ and 
	in the $j$th column and $k$th row, which is $S_{kj}=-S_{jk}= -\sin(\phi_{jk} )$;
		\item boost matrices,  $E^{(n)}_{kj}$ denotes the $n$-dimensional matrix which is defined as:
	all diagonal elements are $1$ save the $k$th and $j$th which are each $c_{jk}=\cos(\epsilon_{jk})$, and
	all nondiagonal elements are zero save the elements in the $k$th column and $j$th row and 
	in the $j$th column and $k$th row, which are $s_{jk}= \imath \sin(\epsilon_{jk} )$.
\end{itemize}
All $\delta_k$, $\phi_{jk}$ and $\epsilon_{jk}$ are real.
For example, when $n=3$,
\begin{eqnarray}
	\Phi^{(3)}_{12}=\left[\begin{array}{ccc}
		C_{12} &  -S_{12} & 0 
		\\
		S_{12} & C_{12} & 0 
		\\
		0 & 0 & 1 
	\end{array}\right], \	E^{(3)}_{12}=\left[\begin{array}{ccc}
		c_{12} & \imath s_{12} & 0 
		\\
		\imath s_{12} & c_{12} & 0 
		\\
		0 & 0 & 1 
	\end{array}\right], \\
	\Phi^{(3)}_{13}=\left[\begin{array}{ccc}
		C_{13} & 0 & -S_{13} 
		\\
		0 & 1 & 0 
		\\
		S_{13} & 0 & C_{13} 
	\end{array}\right],\ 
	E_{13}=\left[\begin{array}{ccc}
		c_{13} & 0 & \imath s_{13} 
		\\
		0 & 1 & 0 
		\\
		\imath s_{13} & 0 & c_{13} 
	\end{array}\right],\\ 
	\Phi^{(3)}_{23}=\left[\begin{array}{ccc}
		1 & 0 & 0 
		\\
		0 & C_{23} & -S_{23} 
		\\
		0 &  S_{23} & C_{23}
	\end{array}\right], \	E_{23}=\left[\begin{array}{ccc}
		1 & 0 & 0 
		\\
		0 & c_{23} & \imath s_{23} 
		\\
		0 & \imath s_{23} & c_{23}
	\end{array}\right].
\end{eqnarray}

%The closed and bounded parameter space is defined by the relations 
%\[
%-\pi\leq \delta_{k}\leq \pi, \quad -\pi/2\leq \phi_{jk}\leq \pi/2, \quad
%-\pi/2\leq \epsilon_{jk}\leq \pi/2.
%\]
Whereas a rotation $\Phi^{(n)}_{kj}V$ preserves the sum of the squares of the components of a column vector $V \in \mathbb{C}^{n}$ in the corresponding $kj$-plane, a boost $E^{(n)}_{kj}V$ preserves the difference of these squares. Of course, both preserve the norm of $V$, $\|V\|^{2} = V^{\dagger}V$, where $\dagger$ denotes the Hermitian conjugate.

Further properties will be used in what follows (for $E^{[n]}$ with arbitrary $n$), as illustrated by the case $n=2$. Let
\begin{eqnarray}
	u=
		\left[\begin{array}{c}
		u_{1}
		\\
	u_{2} 
	\end{array}\right]=
	\left[\begin{array}{c}
		r_{1}+\imath 	i_{2} 
		\\
		r_{2}+\imath 	i_{2} 
	\end{array}\right],\nonumber \\
u'=	\left[\begin{array}{c}
	u'_{1}
	\\
	u'_{2} 
\end{array}\right]=E^{[2]}u=	\left[\begin{array}{cc}
		c &  \imath s
		\\
		\imath s & c
			\end{array}\right]
				\left[\begin{array}{c}
				r_{1}+\imath 	i_{1} 
				\\
				r_{2}+\imath 	i_{2} 
			\end{array}\right]\nonumber \\
			=	\left[\begin{array}{c}
				cr_{1}-si_{2}+\imath \left(	ci_{1}+sr_{2}\right) 
				\\
			cr_{2}-si_{1}+\imath \left(	ci_{2}+sr_{1}\right) 
			\end{array}\right], \label{good_pro1}
	\end{eqnarray} 
	where $r_1$,  $r_2$, $i_1$, and $i_2$ are real.
If $u_{1}^2-u_{2}^2$ and $u'_{1}$ are real, $r_{2}\neq 0$ and $c>0$, then $u'_{2}$ is real and 
	\begin{equation} 
u'=	\begin{bmatrix}
	r_{1}/c
	\\
	r_{2}/c 
\end{bmatrix}.  \label{good_pro2}
	\end{equation} 
	If $u_{1}^2-u_{2}^2$ is real, $i_{1}= 0$,  $r_{2}= 0$, $u'_{1}=0$, and $c>0$, then $u'_{2}$ is imaginary and 
		\begin{equation} 
		u'=	\begin{bmatrix}
		0
			\\
			\imath i_{2}/c 
		\end{bmatrix}.  \label{good_pro3}
	\end{equation} 

%Throughout the following, the superscript $(n)$ denotes an $n$th-order square matrix, $(n,m)$ denotes an $n\times m$ matrix.
Henceforth, the superscript $(n)$ is omitted when it is clear from the context.
I introduce the notation
\begin{eqnarray}
	{\tilde E}_{[j]}= E_{1j}E_{2j}\cdots E_{j-1,j}, \label{tildeE}\\
	{\tilde \Phi}_{[j]}= \Phi_{jn}\Phi_{j,n-1}\cdots \Phi_{j,j+1},	
\end{eqnarray}
where $j<k\leq n$; ${\tilde E}_{[2]}\equiv E_{12}$.

\section{The parametrization of the $n$-dimensional unitary matrix}
	Let $U$ by any $n$-dimensional unitary matrix.  I try to determine the angles $\delta_1$ and $\phi_{1j}$ ($j=2,3,\dots,n$) so that the $n$-dimension matrix 
	\[
	U^{[1]}={\tilde \Phi}_{[1]}^{-1}
U\Delta(-\delta_{1},0,\dots,0)
	\]
	 possesses the following properties:
	\begin{itemize}
		\item $u^{[1]}_{11}$ is real, and $u^{[1]}_{11}\geq 0$;
		\item  all other elements of the first column are imaginary.
	\end{itemize}
I set  $\delta_{1}=\pi/2$ if all $u_{m1}$ ($m=1,\dots,n$) are imaginary, and 
\begin{equation}
	\delta_{1}=\arg \left(\pm \sum_{m=1}^{n}(u_{m1})^{2} \right)/2 \label{delta11}
\end{equation} 
	 otherwise. 
This implies $-\pi/2\leq \delta_{1}\leq \pi/2$.  	Let ${\tilde U}=U\Delta(-\delta_{1},0,\dots,0)$.
		In Eq.~(\ref{delta11}),	the sign is chosen so that
	\begin{equation}
		\sum_{k=1}^{n} \left( {\text Re}({\tilde u}_{k1}) \right)^2> 
	\sum_{k=1}^{n} \left( {\text Im}({\tilde u}_{k1}) \right)^2.  \label{condition1}
\end{equation} 
	If both sides of Eq.~(\ref{condition1}) are equal, the $+$ sign is chosen.

		I now set the real parts of the elements of matrix ${\tilde U}$ in the first column (from the last row up to the second row) to zero by applying successive rotations through angles 
	$-\phi_{1n},-\phi_{1,n-1},\dots, -\phi_{12}$ (Givens rotations). At each step the corresponding angle is defined as $\phi_{1j}=-\pi/2$ if ${\text Re}({\tilde  u}_{11})=0$, and 
	$\phi_{1j}={\arctan}({\text Re}({\tilde u}_{j1})/{\text Re}({\tilde  u}_{11}))$ 
	 otherwise, down to $j=3$.  At $j=2$:  	$\phi_{12}=\text{atan2}({\text Re}({\tilde u}_{j1}),{\text Re}({\tilde  u}_{11}))$.
	Then, recursively,  let
		\[ {\tilde U}:=\Phi_{1 n}^{-1}{\tilde U},\ \dots,\ {\tilde U}:=\Phi_{1 j}^{-1}{\tilde U},\ \dots,\ {\tilde U}:=\Phi_{1 2}^{-1}{\tilde U}. 	\]
  Each rotation alters the value of the first element \({\tilde u}_{11}\), the angles must be defined sequentially based on the updated value of \({\tilde u}_{11}\). This implies $-\pi/2\leq \phi_{1j}< \pi/2$, $j=3,4,\dots, n$, and  $-\pi\leq \phi_{12}< \pi$. The rotations do not change  
	 $\arg \left(\sum_{m=1}^{n}({\tilde u}_{m1})^{2} \right)$ (which is either \(0\) or \(\pi \)); therefore, after the last rotation, ${\tilde u}_{11}\equiv u^{[1]}_{11}>0$ is real, and I get the needed $U^{[1]}$.
	
	If $n=2$, then
	\begin{eqnarray}
	U^{[1]}=
		\left[\begin{array}{cc}
			c  & 	\alpha  			\\
			%%%%%%%%%%%%%%%%%%%%%%%	
			\imath s   & \beta 	
		\end{array}\right],\label{Us2} 
	\end{eqnarray}
	where $c\geq 0$ and $s$ are real.
	Both columns of this matrix are normalized, and orthogonal to each other. 
	 I set $\epsilon_{12}={\text{atan2}}(s,c)$, and apply a boost
		\begin{eqnarray}
		E_{12}^{-1}U^{[1]}=
		\left[\begin{array}{cc}
			1  & 	0  			\\
			%%%%%%%%%%%%%%%%%%%%%%%	
			0   & c'	
		\end{array}\right].\label{RUs2} 
	\end{eqnarray}
	I define 
	$\delta_{2}=\arg(c')$.   
	This implies $-\pi/2\leq \epsilon_{12} < \pi/2$ and $-\pi\leq \delta_{2} < \pi$.
	This yields
	$U=\Phi_{1 2}E_{12} \Delta(\delta_{1},\delta_{2})$.
	
		%%%%%%%%%%%%%%%%%%%%%%%%%% second
		If $n>2$, then I try 
		to determine the angles $\delta_2$ and $\phi_{2j}$ ($j=3,\dots,n$) so that the $n$-dimensional matrix 
		\[
		{\tilde U}^{[2]}=
		{\tilde \Phi}_{[2]}^{-1}
	\Delta(0,-\delta_{2},\dots,0)
		\]
		 possesses the following a property:
		\begin{itemize}
					\item  all elements of the second column bellow $u^{[2]}_{22}$ are imaginary, and 
					$u^{[2]}_{22}$ has a non-zero real part 	${\text Re}\left(u^{[2]}_{22}\right)>0$.
		\end{itemize}
		I set  $\delta_{2}=-\pi/2$ if all $u_{m2}$ ($m=1,\dots,n$) are imaginary, and 
		\begin{equation}
		\delta_{2}=\arg \left( \pm\left(\sum_{m=2}^{n}(u_{m2})^{2}- (u_{12})^{2}\right)\right)/2
		\label{delta2}
		\end{equation}
		 otherwise. 
		 	This implies $-\pi/2\leq \delta_{2}< \pi/2$. 
		 		Let ${\tilde U}:=U^{[1]}\Delta(0,-\delta_{2},\dots,0)$.
	In Eq.~(\ref{delta2}),	the sign is chosen so that
					\begin{equation}
			\sum_{k=2}^{n} \left( {\text Re}({\tilde u}_{k2}) \right)^2> 
			\sum_{k=2}^{n} \left( {\text Im}({\tilde u}_{k2}) \right)^2.
			\label{condition2}
		\end{equation} 
		If both sides of Eq.~(\ref{condition2}) are equal, the $+$ sign is chosen.
		
					I now set the real parts of the elements of matrix ${\tilde U}$ in the second column (from the last row up to the third row) to zero by applying successive rotations through angles 
		$-\phi_{2n},-\phi_{2,n-1},\dots, -\phi_{23}$. 
		At each step the corresponding angle is defined as $\phi_{2j}=-\pi/2$ if ${\text Re}({\tilde  u}_{22})=0$, and 
		$\phi_{2j}={\arctan}({\text Re}({\tilde u}_{j2})/{\text Re}({\tilde  u}_{22}))$ 
		otherwise, down to $j=4$. 
		At $j=3$:  	$\phi_{23}=\text{atan2}({\text Re}({\tilde u}_{32}),{\text Re}({\tilde  u}_{22}))$.
		Then, recursively,  let 
		\[{\tilde U}:=\Phi_{2 n}^{-1}{\tilde U},\ \dots,\ {\tilde U}:=\Phi_{2 j}^{-1}{\tilde U},\ \dots,\ {\tilde U}:=\Phi_{2 3}^{-1}{\tilde U}.\] 
		Each rotation alters the value of the \({\tilde u}_{22}\), the angles must be defined sequentially based on the updated value of \({\tilde u}_{22}\). This implies $-\pi/2\leq \phi_{2j}< \pi/2$, $j=4,5,\dots, n$, and $-\pi\leq \phi_{23}< \pi$. 
		The rotations do not change  the following combination of squares,  
		\begin{equation}
	 \sum_{m=2}^{n}({\tilde u}_{m2})^{2}- ({\tilde u}_{12})^{2}=Const, \label{Const1}
			\end{equation}
		which is real according to Eq.~(\ref{delta2}). %Also, $\text{Re} ({\tilde u}_{22})\neq 0$.
		I get the needed ${\tilde U}^{[2]}$:
		\begin{eqnarray}
		{\tilde U}^{[2]}=
			\left[\begin{array}{cccc}
				r_{1}^{[2]}  & 	 u^{[2]}_{12}  & u^{[2]}_{13} & \dots 			\\
				%%%%%%%%%%%%%%%%%%%%%%%	
				\imath s_{21}^{[2]}   &  u^{[2]}_{22}  & u^{[2]}_{23} 	& \dots \\
				\imath s_{31}^{[2]}   & 	\imath s_{32}^{[2]} & u^{[2]}_{33}	& \dots \\
				\vdots              &   	\vdots & \vdots  & \vdots \\
				\imath s_{n1}^{[2]}   & 		\imath s_{n2}^{[2]} & u^{[2]}_{n3} & \dots
							\end{array}\right]\label{UsX} 
		\end{eqnarray}
			where the  $s_{kl}$ are real (some of them may be zero), $r_{1}^{[2]} $ is positive real by construction in step 1.
		All columns of this unitary matrix are orthonormal.
		I set $\epsilon_{12}={\arctan}(s_{21}^{[2]}/	r_{1}^{[2]})$ (with $-\pi/4\leq\epsilon_{12}\leq\pi/4$,  ensured by Eq.~(\ref{condition1}), which gives $	r_{1}^{[2]}\geq|s_{21}^{[2]}|$) and apply a boost
		\begin{eqnarray}
	E_{12}^{-1}U^{[2]}=
	\left[\begin{array}{cccc}
		r_{1}^{[2]}  & 	 r_{12}^{[2]}   & u^{[2]}_{13} & \dots 			\\
		%%%%%%%%%%%%%%%%%%%%%%%	
	0  &  	r_{2}^{[2]}   & u^{[2]}_{23} 	& \dots \\
		\imath s_{31}^{[2]}   & 	\imath s_{32}^{[2]} & u^{[2]}_{33}	& \dots \\
		\vdots              &   	\vdots & \vdots  & \vdots \\
		\imath s_{n1}^{[2]}   & 		\imath s_{n2}^{[2]} & u^{[2]}_{n3} & \dots
	\end{array}\right]
		.\label{UsssX} 
		\end{eqnarray}		
		Since the columns are orthogonal, $r_{12}^{[2]} $ is real. 
			Again, the boost $E_{12}^{-1}$ does not change  the combination of squares in Eq.~(\ref{Const1}), therefore 
		$	r_{2}^{[2]} $ may be real or imaginary, but Eqs.~(\ref{condition1}) and (\ref{condition2})
		ensure that it is real according to   Eqs.~(\ref{good_pro1}), (\ref{good_pro2}).
		
			If $n=3$, then
	 I set $\epsilon_{13}={\text{atan2}}(s^{[2]}_{31},r^{[2]}_{1})$, and apply a boost
		\begin{eqnarray}
				E_{13}^{-1}E_{12}^{-1}U^{[2]}=
			\left[\begin{array}{ccc}
				1  & 	0 & 0 			\\
				%%%%%%%%%%%%%%%%%%%%%%%	
				0   & 	r_{2}^{[2]} & 	u^{[2]}_{23}   \\
		    	0  & 	\imath  s_{32}' & u'_{33}  
			\end{array}\right],\label{Usss3_1} 
		\end{eqnarray}
			where $s_{32}'$ is real according to Eq.~(\ref{good_pro3}).
						 I set $\epsilon_{23}={\text{atan2}}(s_{32}',r_{2}^{[2]})$, and apply a boost
		\begin{eqnarray}
				E_{23}^{-1}E_{13}^{-1}E_{12}^{-1}U^{[2]}=
			\left[\begin{array}{ccc}
				1  & 	0 & 0 			\\
				%%%%%%%%%%%%%%%%%%%%%%%	
				0   & 1 & 	0  \\
				0  & 0 & \gamma_{3}'' 
			\end{array}\right].\label{Usss3} 
		\end{eqnarray}
			I define $\delta_{3}=\arg(\gamma_{3}'')$, this implies $-\pi\leq \delta_{3} < \pi$.
		This yields
		\[
		U=\Phi^{(3)}_{13}\Phi^{(3)}_{12}\Phi^{(3)}_{23}E^{(3)}_{12} E^{(3)}_{13}E^{(3)}_{23}\Delta^{(3)}(\delta_{1},\delta_{2},\delta_{3}).
		\]
						
	The algorithm is now clear and may be generalized.
		If $n>3$, then, once $\delta_{j}$ has been determined, I have got at a previous step
	%	\begin{widetext}
			\begin{eqnarray}
			 U^{[j]}\equiv  \nonumber
			{\tilde E}_{[j]}^{-1}  {\tilde E}_{[j-1]}^{-1}  \cdots {\tilde E}_{[3]}^{-1}  {\tilde E}_{[2]}^{-1}
				\nonumber\\
				{\tilde \Phi}_{[j]}^{-1} {\tilde \Phi}_{[j-1]}^{-1}\cdots {\tilde \Phi}_{[2]}^{-1} {\tilde \Phi}_{[1]}^{-1}  %\nonumber \\
			U	\Delta(-\delta_{1},-\delta_{2},\dots,-\delta_{j},0,\dots) \nonumber \\=
			\left[\begin{array}{cccccc}
r_{1}^{[j]}& r_{12}^{[j]}  & \dots & r^{[j]}_{1,j} & u^{[j]}_{1,j+1} &	\dots\\
				%%%%%%%%%%%%%%%%%%%%%%%	
0   & r_{2}^{[j]}  	& \dots  & r_{2,j}^{[j]} & u_{2,j+1}^{[j]} &	\dots	 \\
0  & 	0 	& \dots  & r_{3,j}^{[j]} & u_{3,j+1}^{[j]} &	\dots	 \\
	\vdots & 		\vdots & 	\vdots & 	\vdots & 	\vdots & 	\vdots \\
	0 &  0&  \dots & r_{j-1,j}^{[j]}& u_{j-1,j+1}^{[j]}  &  \\
0 &  0&  \dots & r_{j}^{[j]}& u_{j,j+1}^{[j]}  &  \\
%%%%%%%%%%%%%%%%%%%%%%%%%%%%%%%%%%%%%%%%%%%%%%%%%%%%%%%%%%%%%%%%%%%%%%%%%%
\imath s_{j+1,1}^{[j]}   & 	\imath s_{j+1,2}^{[j]} & \dots  &  \imath		s_{j+1,j}^{[j]} & u_{j+1,j+1}^{[j]} &	\dots	\\
%\imath s_{j+2,1}^{[j]} &\imath s_{j+2,2}^{[j]} &\imath s_{j+2,3}^{[j]}	&\dots & \imath s_{j+2,j}^{[j]}  & 	u_{j+2,j+1}^{[j]} &\dots	 \\
\vdots              &   	\vdots  & \vdots  &\vdots & 	\vdots  &	\dots	\\
\imath s_{n1}^{[j]}   & 	\imath s_{n2}^{[j]} & \dots  & \imath s_{nj}^{[j]} & u_{n,j+1}^{[j]} &	\dots	
			\end{array}\right] \nonumber \\ 
			=\left[\begin{array}{cc}
			R^{(j)} & Y 
			\\
			\imath S^{(n-j,j)}  & Z
			\end{array}\right],
			\label{Uj}
		\end{eqnarray}
	%		\end{widetext}
	where $R^{(j)}$ is  a $j$-dimensional upper triangular real matrix, $S^{(n-j,j)}$ is a real matrix,
the	$r_{km}^{[j]}$ and $s_{km}^{[j]}$  are real; some of them may be zero, but the $r_{k}^{[j]}> 0$ ($k=1,\dots, j$) are real.
	
	I try 
	to determine the angles $\delta_{j+1}$ and $\phi_{j+1,k}$ ($k=j+2,\dots,n$) so that the $n$-dimension matrix 
		\begin{eqnarray}
{\tilde U}^{[j]}=	\Phi^{-1}_{[j+1]}
U^{[j]}\Delta(0,\dots,0,-\delta_{j+1},0,\dots)
	\end{eqnarray}
	possesses the following a property:
	\begin{itemize}
		\item  all elements of the $(j+1)$th column bellow ${\tilde u}^{[j]}_{j+1,j+1}$ are imaginary, and 
		${\tilde u}^{[j]}_{j+1,j+1}$ has a non-zero real part 	${\text Re}\left( {\tilde u}^{[j]}_{j+1,j+1}\right)>0$.
	\end{itemize}
	I set  $\delta_{j+1}=\pi/2$ if all $u_{m,j+1}$ ($m=1,\dots,n$) are imaginary, and 
	\begin{equation}
	\delta_{j+1}=\arg \left( \pm\left( \sum_{m=j+1}^{n}(u_{m,j+1})^{2}- \sum_{k=1}^{j} (u_{k,j+1})^{2}\right)\right)/2
	\label{delt_j}
	\end{equation}
	 otherwise. 
	 	 This implies $-\pi/2\leq \delta_{2}\leq \pi/2$.  
	Let 
	\[
	{\tilde U}:=U^{[j]}\Delta(0,\dots,0,-\delta_{j+1},0,\dots,0).
	\]
		In Eq.~(\ref{delt_j}),	the sign is chosen so that
	\begin{equation}
		\sum_{k=j+1}^{n} \left( {\text Re}({\tilde u}_{k,j+1}) \right)^2 > 
		\sum_{k=j+1}^{n} \left( {\text Im}({\tilde u}_{k,j+1}) \right)^2,
		\label{condition4}
	\end{equation} 
	If both sides of Eq.~(\ref{condition4}) are equal, the $+$ sign is chosen.
	
	I  set the real parts of the elements of matrix ${\tilde U}$ in the $(j+1)$th column (from the $n$th row up to the  $(j+2)$th row) to zero by applying successive rotations through angles 
	$-\phi_{j+1,n},-\phi_{j+1,n-1},\dots, -\phi_{j+1,j+2}$. 
	At each step the corresponding angle is defined as $\phi_{j+1,k}=-\pi/2$ if ${\text Re}({\tilde  u}_{j+1,j+1})=0$, and 
	$\phi_{j+1,k}={\arctan}({\text Re}({\tilde u}_{k, j+1})/{\text Re}({\tilde  u}_{j+1,j+1}))$ 
	otherwise, down to $k=j+3$. 
	At $k=j+2$:  	$\phi_{j+1,j+2}=\text{atan2}({\text Re}({\tilde u}_{j+2,j+1}),{\text Re}({\tilde  u}_{j+1,j+1}))$.
	Then, recursively,  let 
	\[
	{\tilde U}:=\Phi_{j+1, n}^{-1}{\tilde U},\ \dots,\ {\tilde U}:=\Phi_{j+1, j+2}^{-1}{\tilde U}. 
	\]
	Each rotation alters the value of the \({\tilde u}_{j+1,j+1}\), the angles must be defined sequentially based on the updated value of \({\tilde u}_{j+1,j+1}\). This implies $-\pi/2\leq \phi_{j+1,k}< \pi/2$, $j=j+3,\dots, n$, and $-\pi\leq \phi_{j+1,j+2}< \pi$. The rotations do not change  
	\[
\sum_{m=j+1}^{n}(u_{m,j+1})^{2}- \sum_{k=1}^{j} (u_{k,j+1})^{2}=Const. 
	\]
	% which is either \(0\) or \(\pi \).
	%Also, $\text{Re} ({\tilde u}_{22})\neq 0$.
	I get 
	%\begin{widetext}
		\begin{eqnarray}
			{\tilde U}^{[j]}  =		  {\tilde \Phi}_{[j+1]}^{-1} 
			U^{[j]}\Delta(0,\dots,0,-\delta_{j+1},0,\dots)  \nonumber
		\\=
			\left[\begin{array}{cccccc}
				r_{1}^{[j]}& r_{12}^{[j]}  & \dots & r^{[j]}_{1,j} & { u}^{[j]}_{1,j+1} &	\dots\\
				%%%%%%%%%%%%%%%%%%%%%%%	
				0   & r_{2}^{[j]} 	& \dots  & r_{2,j}^{[j]} & { u}_{2,j+1}^{[j]} &	\dots	 \\
				%%%%%%%%%%%%%%%%%%%%%%%%%%%%%%%%%%%%%%%%%%%%%%%%%%%%%%%%
				0  & 	0 	& \dots  & r_{3,j}^{[j]} & { u}_{3,j+1}^{[j]} &	\dots	 \\
				%%%%%%%%%%%%%%%%%%%%%%%%%%%%%%%%%%%%%%%%%%%%%%%%%%%%%%%%%%%%%%%%%%
				\vdots & 	\vdots & 	\vdots & 	\vdots & 	\vdots & 	\vdots \\
				0 &  0&   \dots & r_{j-1,j}^{[j]}& { u}_{j-1,j+1}^{[j]}  &  \\
				%%%%%%%%%%%%%%%%%%%%%%%%%%%%%%%%%%%%%%%%%%%%%%%%%%%%%
				0 &  0&  \dots & r_{j}^{[j]}& { u}_{j,j+1}^{[j]}  &  \\
				%%%%%%%%%%%%%%%%%%%%%%%%%%%%%%%%%%%%%%%%%%%%%%%%%%%%%%%%%%%%%%%%%%%%%%%%%%
				\imath s_{j+1,1}^{[j]}   & 	\imath s_{j+1,2}^{[j]} & \dots  &  \imath		s_{j+1,j}^{[j]} & {\tilde r}_{j+1}^{[j]} &	\dots	\\
								\imath s_{j+2,1}^{[j]}   & 	\imath s_{j+2,2}^{[j]}  & \dots  & \imath s_{j+2,j}^{[j]} &
				\imath {\tilde s}_{j+2,j+1}^{[j]}   &	\dots	\\
							\vdots              &   	\vdots   & \vdots  &\vdots & 	\vdots  &	\dots
					\\
				\imath s_{n1}^{[j]}   & 	\imath s_{n2}^{[j]}  & \dots  & \imath s_{nj}^{[j]} &
								\imath {\tilde s}_{n,j+1}^{[j]}   &	\dots	
			\end{array}\right],
			\label{Utilde_j}
		\end{eqnarray}
%	\end{widetext}
	where ${\tilde s}_{k,j+1}^{[j]}$ ($k=j+2,\dots,n$), and  ${\tilde r}_{j+1}^{[j]}\neq 0$ are real.
	I now apply a sequence of boosts zeroing elements of ${\tilde U}^{[j]}$ in the $(j+1)$th row, from first column to the $j$th column. Thus $\epsilon_{k,j+1}={\arctan}\left(s^{[j]}_{j+1,k}/	r_{k}^{[j]}\right)$ with ($-\pi/4\leq\epsilon_{k,j+1}\leq\pi/4$) ($k=1,\dots, j$),  and 
	\[{\tilde U}^{[j]}:=E_{1,j+1}^{-1}	{\tilde U}^{[j]},\ \dots,\ {\tilde U}^{[j]}:=E_{j,j+1}^{-1}	{\tilde U}^{[j]}. \]
	The matrix  ${\tilde \Phi}_{[j+1]}^{-1}$ commutes with ${\tilde E}_{[j]}^{-1}  {\tilde E}_{[j-1]}^{-1}  \cdots {\tilde E}_{[3]}^{-1}  {\tilde E}_{[2]}^{-1}$, so I obtain
	%	\begin{widetext}
		\begin{eqnarray}
			{ U}^{[j+1]}={\tilde E}_{[j+1]}^{-1} % \nonumber\\
			  {\tilde \Phi}_{[j+1]}^{-1} 
			U^{[j]}\Delta(0,\dots,0,-\delta_{j+1},0,\dots)  \nonumber
			\\
			 ={\tilde E}_{[j+1]}^{-1} {\tilde E}_{[j]}^{-1}   \cdots {\tilde E}_{[3]}^{-1}  {\tilde E}_{[2]}^{-1}
			\nonumber\\
			{\tilde \Phi}_{[j+1]}^{-1} {\tilde \Phi}_{[j]}^{-1} \cdots {\tilde \Phi}_{[2]}^{-1} {\tilde \Phi}_{[1]}^{-1}  %\nonumber \\
			U	\Delta(-\delta_{1},-\delta_{2},\dots,-\delta_{j},0,\dots) \nonumber \\=
						\left[\begin{array}{cccccc}
				r_{1}^{[j+1]}& r_{12}^{[j]}   & \dots & r^{[j]}_{1,j} & { u}^{[j]}_{1,j+1} &	\dots\\
				%%%%%%%%%%%%%%%%%%%%%%%	
				0   & r_{2}^{[j+1]}   	& \dots  & r_{2,j}^{[j]} & { u}_{2,j+1}^{[j]} &	\dots	 \\
				%%%%%%%%%%%%%%%%%%%%%%%%%%%%%%%%%%%%%%%%%%%%%%%%%%%%%%%%%%
				0  & 	0	& \dots  & r_{3,j}^{[j]} & { u}_{3,j+1}^{[j]} &	\dots	 \\
				%%%%%%%%%%%%%%%%%%%%%%%%%%%%%%%%%%%%%%%%%%%%%%%%%%%%%%%%%%%%%%%%%%%%%%%%%
				\vdots & 	\vdots & 	\vdots & 	\vdots & 	\vdots & 	\vdots \\
				%%%%%%%%%%%%%%%%%%%%%%%%%%%%%%%%%%%%%%%%%%%%%%%%%%%%%%%%%%%%%%%%%%%%%%
				0 &  0& \dots & r_{j-1,j}^{[j]}& { u}_{j-1,j+1}^{[j]}  &  \\
				%%%%%%%%%%%%%%%%%%%%%%%%%%%%%%%%%%%%%%%%%%%%%%%%%%%%%%%%%%%%%%%%%%%%%%%%
				0 &  0&  \dots & r_{j}^{[j+1]}& { u}_{j,j+1}^{[j]}  &  \\
				%%%%%%%%%%%%%%%%%%%%%%%%%%%%%%%%%%%%%%%%%%%%%%%%%%%%%%%%%%%%%%%%%%%%%%%%%%
	0  & 0 & \dots  & 0 & { r}_{j+1}^{[j+1]} &	\dots	\\
	%%%%%%%%%%%%%%%%%%%%%%%%%%%%%%%%%%%%%%%%%%%%%%%%%%%%%%%%%%%%%
				\imath s_{j+2,1}^{[j]}   & 	\imath s_{j+2,2}^{[j]} & \dots  & \imath s_{j+2,j}^{[j]} &
				\imath {\tilde s}_{j+2,j+1}^{[j]}   &	\dots	\\
				%%%%%%%%%%%%%%%%%%%%%%%%%%%%%%%%%%%%%%%%%%%%%%%%%%%%%%%%%%%%%%%%%%%
	\vdots              &   	\vdots  & \vdots  &\vdots & 	\vdots  &	\dots				\\
	%%%%%%%%%%%%%%%%%%%%%%%%%%%%%%%%%%%%%%%%%%%%%%%%%%%%%%%%%%%%%%%%%%%%%%%%%%%%%%%%%
				\imath s_{n1}^{[j]}   & 	\imath s_{n2}^{[j]} & \dots  & \imath s_{nj}^{[j]} &
				\imath {\tilde s}_{n,j+1}^{[j]}   &	\dots	
			\end{array}\right],
			\label{Utilde_jp1}
		\end{eqnarray}
%	\end{widetext}	
		Since the columns are orthogonal, all ${ u}_{kj}^{[j]}$ ($k=1,\dots, j$) are real. 
		Comparing with Eqs.~(\ref{Uj}) and (\ref{Utilde_j}), one can see which elements have changed at this step. A comparison with Eq.~(\ref{Uj}) then allows the notations to be changed (${ u}_{1,j+1}^{[j]}\equiv { r}_{1,j+1}^{[j+1]}$, etc.), after which one proceeds to the next column.
		
		At the penultimate step, I get
	%		\begin{widetext}
			\begin{eqnarray}
				{\tilde U}^{[n-1]}=
				\left[\begin{array}{cccccc}
					r_{1}^{[n-1]}& r_{12}^{[n-1]}   & \dots & r^{[n-1]}_{1,n-1} & { u}^{[n-1]}_{1,n}\\
					%%%%%%%%%%%%%%%%%%%%%%%	
					0   & r_{2}^{[n-1]}   	& \dots  & r_{2,n-1}^{[n-1]} & { u}_{2,n}^{[n-1]} \\
					%%%%%%%%%%%%%%%%%%%%%%%%%%%%%%%%%%%%%%%%%%%%%%%%%%%%%%%%%%
					0  & 	0 	& \dots  & r_{3,n-1}^{[n-1]} & { u}_{3,n}^{[n-1]} \\
					%%%%%%%%%%%%%%%%%%%%%%%%%%%%%%%%%%%%%%%%%%%%%%%%%%%%%%%%%%%%%%%%%%%%%%%%%
					\vdots & 		\vdots & 	\dots & 	\vdots & 	\vdots \\
					%%%%%%%%%%%%%%%%%%%%%%%%%%%%%%%%%%%%%%%%%%%%%%%%%%%%%%%%%%%%%%%%%%%%%%
					0 &  0&  \dots & r_{n-2,n-1}^{[j]}& { u}_{n-2,n}^{[j]}  \\
									%%%%%%%%%%%%%%%%%%%%%%%%%%%%%%%%%%%%%%%%%%%%%%%%%%%%%%%%%%%%%%%%%%%%%%%%%%
					0  & 0 &  \dots  & { r}_{n-1}^{[n-1]} & { u}_{n-1,n}^{[n-1]} 	\\
					%%%%%%%%%%%%%%%%%%%%%%%%%%%%%%%%%%%%%%%%%%%%%%%%%%%%%%%%%%%%%
					\imath s_{n1}^{[n-1]}   & 	\imath s_{n2}^{[n-1]} & \dots  & \imath s_{n,n-1}^{[n-1]} &
					{ u}_{n,n}^{[n-1]} 	
				\end{array}\right],
				\label{Ulast}
			\end{eqnarray}
	%	\end{widetext}	
		Matrix $	{\tilde U}^{[n-1]}$ is unitary; therefore, just as for $n=3$, a boost 
			 	$E_{1n}^{-1}$ with $\epsilon_{1n}={\text{atan2}}(s_{n1},	r_{1}^{[n-1]})$ will zero not only 
			 	the $(n,1)$ element but all elements of the first row except the 	$(1,1)$ element.
			 	Applying then consequently boosts $E_{kn}^{-1}$ with $\epsilon_{kn}={\text{atan2}}({\tilde s}_{nk},	r_{k}^{[n-1]})$ ($k=1,\dots,n-1$) to the matrix ${ U}^{[n-1]}$ (at each step ${\tilde s}_{nk}$ must be redefined for all $k=1,\dots,n-1$), one obtains a diagonal matrix whose diagonal elements are all equal to unity except the $(n,n)$ element. That last element (let it be $u^{[n]}$) gives a last angle $\delta_{n}=\arg(u^{[n]})$, $-\pi\leq \delta_{n} < \pi$.
			This yields the parametrization of the unitary matrix, 
	%	\begin{widetext}	
		\begin{eqnarray}
				U={\tilde \Phi}_{[1]}{\tilde \Phi}_{[2]}\cdots{\tilde \Phi}_{[n-1]}
				{\tilde E}_{[2]}{\tilde E}_{[3]}\cdots	 	{\tilde E}_{[n]} \Delta(\delta_{1},\dots,\delta_{n}).
			\end{eqnarray}
%		\end{widetext}	
	%
		\section{Parametrization of the $S$-matrix}
Using a standard algorithm, a unitary symmetric $n$-dimensional $S$ matrix can be diagonalized by a real rotation matrix $A$:
\begin{equation}
S=A^{T}D^{2}A=(DA)^{T}DA,
	\label{S_Blatt1}
\end{equation}
where $D$ is a diagonal matrix
\begin{equation}
	D=
	diag\left({\mathrm e}^{\imath \delta_1},{\mathrm e}^{\imath \delta_2},\dots,{\mathrm e}^{\imath \delta_n}\right);
	\label{D_Blatt1}
\end{equation}
with eigenphases $\delta_{k}$, $k=1,\dots,n$.
%For any matrix $B$ let $a_{jk}$ be the element in the $j$th row and $k$th column of $B$, $j,k=1,\dots,n$; $b_{(k)}$ denotes $k$th column of $B=\left[b_{(1)},\dots,b_{(n)}\right]$

I define the bar parametrization as
\begin{equation}
	S=\Delta {\tilde E}^{T}{\tilde E} \Delta =({\tilde E} \Delta)^{T}{\tilde E} \Delta,
	\label{S_Bar1}
\end{equation}
where
\begin{equation}
	\Delta =	
		diag\left({\mathrm e}^{\imath \bar{\delta}_1},{\mathrm e}^{\imath \bar{\delta}_2},\dots,{\mathrm e}^{\imath \bar{\delta}_n}\right),
			\label{D_Bar1}
\end{equation}
and 
\begin{equation}
	{\tilde E}={\tilde E}_{[2]}{\tilde E}_{[2]}\cdots {\tilde E}_{[n]}.
	\label{V_Bar1}
\end{equation}
The ${\tilde E}_{[j]}$ are defined by the $\epsilon_{kj}$ ($k=1,\dots,j-1$) according to  Eq.~(\ref{tildeE}).
It remains to find the bar phases $\bar{\delta}_{k}$ and mixing angles $\epsilon_{ji}$ required to represent the matrix $S$ in the form of Eq.~(\ref{S_Bar1}).
The relation
 \begin{equation}
 \sum_{k=1}^{n} { \delta}_{k} = \sum_{k=1}^{n} {\bar \delta}_{k}\ \text{ modulo} \ \pi \label{sum_delta}
\end{equation}
  may be used in the extraction procedure, though it is not necessary.

Eqs.~(\ref{S_Blatt1}) and (\ref{S_Bar1}) are Takagi factorizations of the same unitary symmetric matrix. Clearly,
\begin{equation}
		DA=W	{\tilde E} \Delta,
	\label{WDA}
\end{equation}
where $W$ is some rotation-reflection matrix,  so that $W^{T}W=I$, with $I$ the identity matrix. 
Moreover, the algorithm presented in Sect.~III allows one to find the required factorization
\begin{equation}
	DA ={\tilde \Phi}_{[1]}{\tilde \Phi}_{[2]}\cdots{\tilde \Phi}_{[n-1]}
	{\tilde E} \Delta. \label{DA=PhiED}
\end{equation}
This yields the needed parametrization
	\begin{eqnarray}
	S= \Delta({\bar \delta}_{1},\dots,{\bar \delta}_{n}) {\tilde E}_{[n]}{\tilde E}_{[n-1]}\cdots{\tilde E}_{[3]}
\nonumber\\
	{\tilde E}_{[2]}^2{\tilde E}_{[3]}\cdots	 	{\tilde E}_{[n]} \Delta({\bar \delta}_{1},\dots,{\bar \delta}_{n}).
\end{eqnarray}

Equation~(\ref{DA=PhiED}) defines the matrix ${\tilde E} \Delta$ up to a certain reflection. This reflection can be determined at the final step of the algorithm described in Sect.~III (provided $|\bar{\delta}_{n}|>\pi/2$) and its effect corrected by pre-multiplying $DA$ by a reflection matrix
$R=diag\left(1,1,\dots,1,-1\right)$. This redefines the interval for $\delta_{n}$ as  $-\pi/2\leq \delta_{n} < \pi/2$.

Let us consider the case $n=3$ in more detail. In this case,
\begin{widetext}
	\begin{eqnarray}
		{\tilde E} \Delta ={\tilde E}_{[2]}{\tilde E}_{[3]} \Delta({\bar \delta}_{1},\dots,{\bar \delta}_{n})\nonumber\\
		\equiv E_{12}E_{13}E_{23}\Delta({\bar \delta}_{1},\dots,{\bar \delta}_{n})
		=
		\left[\begin{array}{ccc}
			c_{12}c_{13}{\mathrm e}^{\imath {\bar \delta}_1}   & \left(\imath 	s_{12}	c_{23} - c_{12} s_{13} s_{23} \right) {\mathrm e}^{\imath 	{\bar \delta}_2} & \left(-s_{12} s_{23}+\imath c_{12} s_{13} c_{23}\right)		{\mathrm e}^{\imath {\bar \delta}_3} 	\\
			%%%%%%%%%%%%%%%%%%%%%%%	
			\imath s_{12} c_{13} {\mathrm e}^{\imath {\bar  \delta}_1} & \left(c_{12}c_{23}-\imath s_{12} s_{13} s_{23}\right) 	{\mathrm e}^{\imath {\bar \delta}_2} & \left(\imath c_{12} s_{23}-s_{12} s_{13} c_{23}\right) 	{\mathrm e}^{\imath {\bar \delta}_3} 	\\
			%%%%%%%%%%%%%%%%%%%%%%%%%%
			\imath s_{13} {\mathrm e}^{\imath {\bar \delta}_1} & \imath c_{13} s_{23} {\mathrm e}^{\imath {\bar \delta}_2} & c_{13} c_{23} {\mathrm e}^{\imath {\bar \delta}_3} 
		\end{array}\right].\label{VD_form}
	\end{eqnarray}
\end{widetext}
The algorithm of Sect.~III finds this matrix up to a reflection-matrix factor on the left.
The degeneracy of ${\bar \delta}_1$ is modulo $\pi/2$ as is clear from Eq.~(\ref{delta11}) that is equivalent to
\begin{equation}
{\bar \delta}_{1}=\arg \left(\pm S_{11} \right)/2. \label{delta11_2}
\end{equation} 
Then $	{\bar \delta}_{1}$ is fixed by some condition, for example by Eq.~(\ref{condition1});
the degeneracy of ${\bar \delta}_2$ is again modulo $\pi/2$ as is clear from Eq.~(\ref{delta2}).
That equation also may be rewritten in terms of the $S$-matrix  elements, but in a more cumbersome way than Eq.~(\ref{delta11_2}). ${\bar \delta}_2$ is then fixed by some condition, for example by Eq.~(\ref{condition2}). The remaining $	{\bar \delta}_{3}$ is then obviously fixed by Eq.~(\ref{sum_delta}) and may be fixed properly by pre-multiplying $DA$ by a reflection matrix
$R=diag\left(1,1,-1\right)$. Then all $\bar{\delta}_{i}$ are fixed; the angles $\epsilon_{12}$, $\epsilon_{13}$, and $\epsilon_{23}$ are each degenerate modulo $\pi$.

\section{Conclusions}
I have presented a parametrization of the symmetric unitary $S$-matrix that is a generalization of the Stapp parametrization to an arbitrary number of coupled channels. I suppose that the presented parametrization will be useful for the analysis of $NN$ scattering above the inelastic threshold and for similar problems. Moreover, it will undoubtedly be used in the continuation of my work on solving the inverse scattering problem~\cite{My2025,My2026}.
The parametrization of a unitary matrix obtained in solving the main problem may prove useful both in quantum scattering theory and in other areas of physics.

The code for the parametrization of the unitary matrix and for the parametrization of the unitary symmetric $S$ matrix is available in the Maple programming language as a text file~\cite{Data2026}.
%
%&&&&&&&&&&&&&&&&&&&&&&&&&&&&&&&&&&&&&&&
\section*{Acknowledgements}
During the preparation of this work, the author used DeepSeek AI for language editing and clarity improvements. After using this tool/service, the author reviewed and edited the content as needed and takes full responsibility for the content of the publication.
\section*{Data availability}
The data that support the findings of this article are openly
available \cite{Data2026}.
%
%\cite{Blatt1952,Stapp1957,MacGregor,Hoshizaki} %\cite{Bryan1981,Bryan1984,Arndt1982,Arndt1983,Sprung1982,Sprung1985}

%
\end{document}